\pdfoutput=1
\documentclass[10pt,conference]{IEEEtran}
\usepackage[usenames,dvipsnames,table]{xcolor}
\usepackage{amsmath,amssymb,amsfonts,amsthm}
\usepackage[pdftex]{graphicx}
\usepackage{tabularx}

\usepackage[T1]{fontenc}
\usepackage[utf8]{inputenc}

\usepackage{xfrac}
\usepackage{physics}

\usepackage{adjustbox}
\usepackage{csquotes}
\usepackage{nicefrac}
\usepackage{csvsimple}

\usepackage[textsize=tiny]{todonotes}
\usepackage{booktabs}
\usepackage[hyphens,spaces,obeyspaces]{url}
\usepackage[draft]{hyperref}
\usepackage{tikz}
\usetikzlibrary{quantikz}
\usepackage{pgfplots}
\usepackage{graphicx}

\usetikzlibrary{shapes.geometric, positioning, quantikz}

\makeatletter
\let\MYcaption\@makecaption
\makeatother
\usepackage[font=footnotesize]{subcaption}
\makeatletter
\let\@makecaption\MYcaption
\makeatother

\usepackage[style=ieee, maxnames=6, minnames=1, date=year, doi=false,isbn=false,backend=biber]{biblatex}
\usepackage[binary-units=true, detect-all=true]{siunitx}
\usepackage{etoolbox}
\robustify\bfseries

\usepackage{nicematrix}
\usepackage{flushend}

\newtheorem{example}{Example}

\tikzset{%
	font={\footnotesize},
	terminal/.style={draw,rectangle,inner sep=2pt,font=\footnotesize,very thick},
	vertex/.style={draw,circle,inner sep=0pt,minimum width=0.5cm,minimum height=0.5cm},
	define color/.code={\definecolor{hsb#1}{Hsb}{#1, 1, 0.75}},				
    medge/.style n args={3}{
		line width={#1pt},
		define color={#2},
		draw=hsb#2,
		out=#3, 
		in=90
	},
    edge/.style 2 args={
		line width={#1pt},
		define color={#2},
		draw=hsb#2
	},
	edge0/.style 2 args={
		line width={#1pt},
		define color={#2},
		draw=hsb#2,
		out=-130, 
		in=90
	},
	edge1/.style 2 args={
		line width={#1pt},
		define color={#2},
		draw=hsb#2,
		out=-50, 
		in=90
	},
	zerostub/.style={
		inner sep=0, 
		minimum size=3pt, 
		circle, 
		fill=black
	}
}

\begin{document}

\title{Automatic Implementation and Evaluation\\of \mbox{Error-Correcting} Codes for Quantum Computing\\[.2em]\Large{An Open-Source Framework for Quantum Error Correction}}

\author{
	\IEEEauthorblockN{Thomas Grurl\IEEEauthorrefmark{1}\IEEEauthorrefmark{2}\hspace*{1.5cm}Christoph Pichler\IEEEauthorrefmark{2}\hspace*{1.5cm}Jürgen Fuß\IEEEauthorrefmark{1}\hspace*{1.5cm}Robert Wille\IEEEauthorrefmark{3}\IEEEauthorrefmark{4}}
	\IEEEauthorblockA{\IEEEauthorrefmark{1}University of Applied Sciences Upper Austria, Austria}
	\IEEEauthorblockA{\IEEEauthorrefmark{2}Johannes Kepler University Linz, Austria}
	\IEEEauthorblockA{\IEEEauthorrefmark{3}Chair for Design Automation, Technical University of Munich, Germany}
	\IEEEauthorblockA{\IEEEauthorrefmark{4}Software Competence Center Hagenberg GmbH (SCCH), Austria}
	\IEEEauthorblockA{\href{mailto:thomas.grurl@fh-hagenberg.at}{thomas.grurl@fh-hagenberg.at}\hspace{0.5cm}
	\href{mailto:christoph.pichler@jku.at}{christoph.pichler@jku.at}\hspace{0.5cm}
	\href{mailto:juergen.fuss@fh-hagenberg.at}{juergen.fuss@fh-hagenberg.at}\hspace{0.5cm} 
	\href{mailto:robert.wille@tum.de}{robert.wille@tum.de}\\
	\href{www.cda.cit.tum.de/research/quantum/}{www.cda.cit.tum.de/research/quantum/}\\ \mbox{ }
	}
	\vspace*{-2.1em}
}

\maketitle

\begin{abstract}
Due to the fragility of quantum mechanical effects, real quantum computers are plagued by frequent noise effects that cause errors during computations. Quantum \mbox{\mbox{error-correcting}} codes address this problem by providing means to identify and correct corresponding errors. However, most of the research on quantum error correction is theoretical or has been evaluated for specific hardware models only. Moreover, the development of corresponding codes and the evaluation of whether they indeed solve the problem for a particular hardware model, still often rests on tedious trial-and-error thus far. In this work, we propose an \mbox{open-source} framework that supports engineers and researchers in these tasks by automatically applying \mbox{error-correcting} codes for a given application followed by an automatic noise-aware quantum circuit simulation. Case studies showcase that this allows for a substantially more efficient implementation and evaluation of \mbox{error-correcting} codes.

\end{abstract}

\section{Introduction}
\label{sec:introduction}

Quantum computers can solve specific problems substantially faster than classical computers. 
Examples for such problems include algorithms that have been found in the areas of chemistry~\cite{McArdle2020}, machine learning~\cite{Biamonte2017}, biology~\cite{Cordier2021}, and finance~\cite{Egger2021}.
They achieve this by exploiting quantum mechanical effects during their computations, such as superposition (i.e., a qubit can be in a combination of 0 and 1) and entanglement (i.e., measurement outcomes of individual qubits are correlated). 
However, the same properties that make quantum computing so powerful also make it extremely prone to errors. For example, while it is a key feature of quantum bits that they can be in a superposition of 0 and 1, this also makes them fragile against operational or environmental noise. Accordingly, it is believed that (quantum) error correction is going to be an essential part of future quantum hardware~\cite{Preskill2018quantumcomputingin}. 

But error correction is not easy. That it is even possible in the quantum realm is not straightforward. 
The first algorithm demonstrating this was published in 1995 by Peter Shor~\cite{shor1995}. 
Since then %
there have been considerable developments in the research on quantum error correction. Today, there are multiple types of quantum \mbox{error-correcting} schemes available such as CSS codes~\cite{shor1995,Steane1996,Calderbank1996}, stabilizer codes~\cite{Gottesman1996}, or surface codes~\cite{Bravyi1998,Dennis2002}. Additionally, there have also been developments in adaptive \mbox{error-correcting} codes optimized for specific quantum hardware~\cite{Fletcher2008,Robertson2017}.

However, despite this progress, most of the corresponding work in this domain still heavily relies on manual labor and/or is based on theoretical results only. To the best of our knowledge, methods (or implementations thereof) that take a quantum circuit and \emph{automatically} extend it with a correspondingly chosen \mbox{error-correcting} scheme (similar to a compiler that translates a given quantum algorithm to a corresponding hardware-applicable realization) do not exist, yet. This may be caused by the fact that, thus far, the focus clearly (and understandably) is on the development of %
concepts for \mbox{error-correcting} codes and proof-of-concept implementations on selected hardware models only. 
Additionally, the development of such a framework is not straightforward. For example, the decoding and correction steps are exponentially hard~\cite{Hsieh2011} and, while this is addressed in recent work (such as~\cite{aspdac_2023_4}), still remain complex. 
Furthermore, applying quantum operations to qubits protected by an \mbox{error-correcting} code is often not directly possible but requires specific routines (as shown by the no-go theorem~\cite{eastin2009}).

As a consequence, evaluations on whether an \mbox{error-correcting} code is useful in different scenarios are cumbersome and, hence, often done with rather selected use cases only. 
Because of this, comprehensive case studies in which the usefulness of \mbox{error-correcting} codes are evaluated with respect to different circuits, how they are configured, and on what hardware model they are applied, do not exist.

In this work we address this issue: We propose an \mbox{open-source} framework\footnote{The framework is available at \href{www.github.com/cda-tum/qecc}{www.github.com/cda-tum/qecc}}
that supports engineers and researchers in the task of evaluating \mbox{error-correcting} codes. The framework allows to \emph{automatically} apply error correction schemes to 
a given quantum circuit, followed by an automatic noise-aware quantum circuit simulation. To this end, 
we implemented different \mbox{error-correcting} codes
and utilized existing methods for quantum circuit simulation.  %
In this way, a framework is created that allows for \mbox{error-correcting} codes to be easily analyzed---with minimal manual effort. 
The proposed framework is implemented in such a modular way that it can be readily extended for new \mbox{error-correcting} codes or different simulation styles. Case studies showcase that the proposed framework allows for efficient evaluations of \mbox{error-correcting} codes depending on varying properties.

The remainder of the paper is organized as follows: In Section~\ref{sec:background}, we review the basics of quantum computing and noise in quantum devices. Afterward, in Section~\ref{sec:motivation}, we motivate the proposed framework by revisiting the basics of quantum error correction and illustrating the current problems. Section~\ref{sec:framework} presents the proposed framework, followed by Section~\ref{sec:evaluation} demonstrating the application and usefulness of the framework. Finally, Section~\ref{sec:conclusions} concludes the paper.

\section{Preliminaries}
\label{sec:background}

In this section, we review the basics of quantum computing as well as noise in quantum computing. We refer the interested reader to~\cite{NC:2000} for a thorough introduction to this topic. 

\subsection{Quantum Computing}
\label{sec:quantum_computing}

The basic unit of information in the quantum world is called a \emph{quantum bit} or \emph{qubit}. Like a classical bit, a qubit can assume the states 0~and~1, but, additionally, it can also assume an arbitrary combination of those \emph{basis states}. More generally---using Dirac notation---the state of a qubit $\ket{\psi}$ can be written as 
\mbox{$\ket{\psi} = \alpha_0 \cdot \ket{0} + \alpha_1 \cdot \ket{1}$} with $\alpha_0, \alpha_1 \in \mathbb{C}$ and the normalization constraint \mbox{$\abs{\alpha_0}^2 + \abs{\alpha_1}^2 = 1$}. The values $\alpha_0, \alpha_1$ are called \emph{amplitudes} and specify how strongly the qubit relates to each of the basis state $\ket{0}$ and $\ket{1}$. Measuring a qubit collapses it to a basis state. More precisely, with probability $\abs{\alpha_0}^2$ the state $\ket{\psi}$ collapses to $\ket{0}$, and with probability $\abs{\alpha_1}^2$ it collapses to $\ket{1}$. Only the basis state is returned by the measurement, the amplitudes cannot be directly measured. 
The state description can be extended for \mbox{multi-qubit} systems, e.g.,~a two-qubit state $\ket{\phi}$ is fully characterized by $2^2=4$ amplitudes  and, hence,
\mbox{$\ket{\phi} = \alpha_{00} \cdot \ket{00} + \alpha_{01} \cdot \ket{01} + \alpha_{10} \cdot \ket{10} + \alpha_{11} \cdot \ket{11}$}. 

Having this formalism to describe quantum states, the next question becomes how to modify them. This is done by using so-called quantum operations. Important single-qubit operations are 
the Hadamard ($\operatorname{H}$) operation, which transforms a qubit from a basis state into a superposition (\mbox{$\operatorname{H}\ket{\psi} = \nicefrac{\alpha_0}{\sqrt{2}} \cdot (\ket{0} + \ket{1}) + \nicefrac{\alpha_1}{\sqrt{2}} \cdot (\ket{0} - \ket{1})$}), the $\operatorname{X}$~operation, which is the quantum equivalent of flipping a bit (\mbox{$\operatorname{X}\ket{\psi} = \alpha_0 \cdot \ket{1} +  \alpha_1 \cdot \ket{0}$}), and the $\operatorname{Z}$~operation, which flips the phase of a qubit (\mbox{$\operatorname{Z}\ket{\psi} = \alpha_0 \cdot \ket{0} -  \alpha_1 \cdot \ket{1}$}).
Two-qubit operations are also possible. An essential two-qubit operation is the \emph{Controlled NOT} ($\operatorname{CX}$), which flips the target qubit \emph{iff} the control qubit is set to one (\mbox{$\operatorname{CX}\ket{\phi} = \alpha_{00} \cdot \ket{00} + \alpha_{01} \cdot \ket{01} + \alpha_{10} \cdot \ket{11} + \alpha_{11} \cdot \ket{10}$}).

Quantum circuits are often illustrated as circuit diagrams. Diagrams consist of one or more vertical lines which represent qubits. Per convention, the top line represents qubit 0 and the remaining qubits are labeled sequentially. The lines are interrupted by operations that are applied to the respective qubit, which is usually indicated by labeled rectangles. In a quantum circuit, time flows from left to right, i.e., the leftmost operation on a qubit wire is applied first, and then the second left-most operation is applied, and so on.

\begin{figure}
	\center
	\begin{quantikz}
	\lstick{$\ket{0}$} 		 & \gate{H} &  \ctrl{1} &  \ctrl{2} & \meter{} & \cw \\
	\lstick{$\ket{0}$}       & \qw      &  \targ{}  &  \qw      & \meter{} & \cw \\
	\lstick{$\ket{0}$}       & \qw      &  \qw      &  \targ{}  & \meter{} & \cw \\
	\end{quantikz}
	\vspace*{-5mm}
	\caption{Quantum circuit constructing the maximally entangled state}
	\label{fig:ghz_state_example}
	\vspace*{-5mm}
\end{figure}
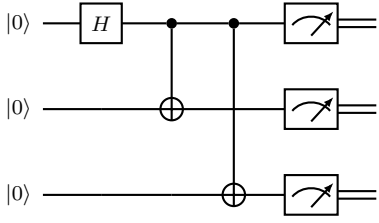

\begin{example}
\label{exp:quantum_circuit}
In Fig.~\ref{fig:ghz_state_example}, a quantum circuit is depicted that constructs a maximally entangled state over three qubits. The circuit consists of three qubits which are all initialized in basis state $\ket{0}$. Thus, the initial state is $\ket{000}$. Applying the $\operatorname{H}$~operation to qubit 0 (i.e., the topmost line) changes the state to
\vspace*{-1mm}
\begin{align*}
\operatorname{H} \ket{000} = \nicefrac{1}{\sqrt{2}} \cdot (\ket{000} + \ket{100})
\end{align*}
The next operations are two $\operatorname{CX}$ gates. The $\bullet$ symbol marks the control qubit and the $\oplus$ specifies the target qubit. These yield 
\vspace*{-1mm}
\begin{align*}
\operatorname{CX} \nicefrac{1}{\sqrt{2}} \cdot (\ket{000} + \ket{100}) = \nicefrac{1}{\sqrt{2}} \cdot (\ket{000} + \ket{110})\text{ and } \\ 
\operatorname{CX} \nicefrac{1}{\sqrt{2}} \cdot (\ket{000} + \ket{110}) = \nicefrac{1}{\sqrt{2}} \cdot (\ket{000} + \ket{111}).\;\;\;\;\;\;\;
\end{align*}
Finally, all qubits are measured, indicated by the last operation on each lane. After the measurement, the qubits only contain classical information, which is indicated by the double wire. The final state is either $\ket{000}$ or $\ket{111}$, each with a probability of~${\sfrac{1}{2}}$. Note that, in this case, measuring one qubit affects the other qubits as well. This is an essential concept of quantum computing called \emph{entanglement}.
\end{example}

\subsection{Noise and Errors in Quantum Computing}
\label{sec:noise_in_quantum_computing}

Example~\ref{exp:quantum_circuit} illustrates how \emph{ideal} quantum computers behave. Unfortunately, \emph{real} quantum computers are plagued by noise effects, which degrade computations and introduce errors. 
Although the error probability of quantum systems is constantly reduced, noise will always be a part of quantum computing~\cite{Preskill2018quantumcomputingin}. 
In the literature, errors resulting from these noise effects are often distinguished between \emph{operational errors} and \emph{coherence errors}~\cite{Tannu2018}.

\emph{Operational errors} are introduced whenever an operation is applied to a qubit~\cite{Tannu2018}. They occur since quantum computers are mechanical apparatuses, susceptible to (tiny) errors. Hence, whenever an operation is applied to a qubit, it may be not executed at all, or in a (slightly) modified fashion. Operational errors are highly specific to each quantum computer, therefore they are often simulated using a depolarization error~\cite{qiskit}. This error simulates that a qubit is set to a completely random state~\cite{NC:2000}.

\emph{Coherence errors} occur because of the fragile nature of qubits. This fragility leads to the problem that qubits can hold information only for a limited amount of time. Two types of coherence errors are usually distinguished~\cite{Tannu2018}: 
\begin{itemize}
\item Qubits can lose or gain energy from the environment, i.e., a qubit in a \mbox{high-energy} state ($\ket{1}$) tends to relax into a low-energy state ($\ket{0}$) and vice-versa. This is called \emph{amplitude damping} or \emph{T1~error}. 

\item When qubits interact with the environment it can happen that a phase flip effect occurs. This leads to a \emph{phase-flip} or \emph{T2~error}.
\end{itemize}

\section{General Idea and Motivation}
\label{sec:motivation}

As reviewed in Section~\ref{sec:noise_in_quantum_computing}, quantum computers are plagued by noise, which drastically limits their usefulness in the real world. Quantum error correction tackles this problem, which makes it an essential part of building scalable and resilient quantum hardware. 
This section reviews the main ideas of the corresponding concepts and provides a motivation why an easy-to-use framework 
for their %
evaluation is needed. 

\subsection{Quantum Error Correction}
\label{subsec:quantum_error_correction}

To illustrate quantum error correction, we first revisit classical error correction, as it serves the same purpose. In the classical world, error correction is achieved by encoding information beyond its theoretical minimum. The redundant information is then used to identify and (if possible) correct errors. To illustrate this, consider the following example:

\begin{example}
\label{exp:classical_ecc}
Suppose a sender wants to transfer a single (classical) bit 0 to a receiver at a different location. The transmission channel is prone to bit-flip errors, which is why both sender and receiver agree to protect the bit, using the three-bit repetition code.
For this code, each bit of information is encoded by tripling it, i.e., $0 \rightarrow 000$ and $1 \rightarrow 111$. Hence, the \mbox{one-bit} message 0 is encoded into the codeword `000' before being transmitted to the receiver. During the transmission a bit-flip error occurs (distorting the message, e.g., to 001), the receiver could, through majority voting, still infer that the sent message most likely was 000 and could therefore correctly restore the original message 0.
\end{example}

In the quantum world, error correction is done similarly, but some properties of quantum mechanics make it more complex: 
A key difference between the quantum and the classical world is that measurements in the quantum world affect the observed system.
So, while the classical system can be measured without risk of compromising the encoded information, special care must be taken in quantum error correction as to not destroy information by measuring it. This is worsened by the \mbox{no-cloning} theorem~\cite{Wootters1982Single}, which asserts that it is not possible to clone (i.e., copy) arbitrary quantum states.
It is therefore not possible to simply clone qubits before measuring them to keep their information intact. 

Instead, in order to add redundancy in the quantum world, we expand the Hilbert space, in which the information is encoded---effectively distributing the information of a single qubit among more qubits~\cite{shor1995}.
We illustrate this using the three-qubit bit-flip code, which allows detection and correction of single-qubit bit-flip errors. Using this code, a single qubit is encoded by entangling it with two ancillary qubits (this can be achieved by two CX operations, similar to Example~\ref{exp:quantum_circuit}), i.e., 
\begin{equation}
\ket{\psi} = \alpha_0 \cdot \ket{0} + \alpha_1 \cdot \ket{1} \rightarrow \ket{\psi_L} = \alpha_0 \cdot \ket{000} + \alpha_1 \cdot \ket{111}.
\label{eq:ecc3_encoding}
\end{equation}
After encoding the information of state $\ket{\psi}$, it is distributed among the 3-qubit state $\ket{\psi_L}$.

More precisely, the information of $\ket{\psi}$ (encoded into the 2-dimensional Hilbert space $\operatorname{span}\{\ket{0},\ket{1}\}$) is now encoded into the 8-dimensional Hilbert space $\operatorname{span}\{\ket{000},\ket{001}, \hdots, \ket{111}\}$. %
This 8-dimensional Hilbert space can be split into 4 subspaces $\mathcal{C} = \operatorname{span}\{\ket{000},\ket{111}\}$, $\mathcal{F}_1 = \operatorname{span}\{\ket{001},\ket{110}\}$, $\mathcal{F}_2 = \operatorname{span}\{\ket{010},\ket{101}\}$, and $\mathcal{F}_3 = \operatorname{span}\{\ket{100},\ket{011}\}$, where the subspace  $\mathcal{C}$ represents the \emph{logical code space} (indicating that the system is in a valid state) and the subspaces $\mathcal{F}_1, \mathcal{F}_2$, as well as $\mathcal{F}_3$ represent 
\emph{logical error spaces} (indicating %
a bit-flip error). %
That is, $\mathcal{F}_1$ indicates that a bit-flip error in the first qubit occurred, $\mathcal{F}_2$ indicates a bit-flip at the second qubit, and $\mathcal{F}_3$ indicates a bit-flip in the third qubit. By using a special kind of measurement it is possible to infer which qubits are equal (we provide an example realization of such an indirect measurement later in Section~\ref{sec:compiler}). With this knowledge, it is possible to infer in which of the subspace $\mathcal{C}, \mathcal{F}_1,\mathcal{F}_2,\mathcal{F}_3$ $\ket{\psi_L}$ resides. Note that this measurement does not produce any information whether the encoded qubit is 0 or 1 and, therefore, does not change the encoded qubit.

\begin{example}
\label{exp:shor3}
Suppose a sender wants to transfer the single qubit $\ket{\psi^\prime} = \sqrt{\nicefrac{1}{3}} \cdot \ket{0} + \sqrt{\nicefrac{2}{3}} \cdot \ket{1}$ to a receiver at a different location. The transmission channel is prone to bit-flip errors, so sender and receiver agree to protect the qubit using the three-qubit bit-flip code (presented above). The sender encodes the qubit $\ket{\psi^\prime} = \sqrt{\nicefrac{1}{3}} \cdot \ket{0} + \sqrt{\nicefrac{2}{3}} \cdot \ket{1} \rightarrow \ket{\psi^\prime_L} = \sqrt{\nicefrac{1}{3}} \cdot \ket{000} + \sqrt{\nicefrac{2}{3}} \cdot \ket{111}$ and sends $\ket{\psi^\prime_L}$ to the receiver.

During the transmission the first qubit flips---leaving it in state $\sqrt{\nicefrac{1}{3}} \cdot \ket{100} + \sqrt{\nicefrac{2}{3}} \cdot \ket{011}$. 
Before using the qubit, the receiver measures if all physical qubits making up $\ket{\psi^\prime_L}$ are equal. The measurement shows that the first qubit is different from the other two qubits and, hence, the receiver infers that $\ket{\psi^\prime_L}$ resides in the subspace $\mathcal{F}_3$, indicating that a bit-flip error has occurred in the first qubit. The receiver therefore applies an $\operatorname{X}$ operation to the first qubit of $\ket{\psi^\prime_L}$---restoring it back to $\ket{\psi_L} = \sqrt{\nicefrac{1}{3}} \cdot \ket{000} + \sqrt{\nicefrac{2}{3}} \cdot \ket{111}$. Afterwards, the receiver can decode $\ket{\psi^\prime_L}$ to $\ket{\psi^\prime}$ and use it for future computations.
\end{example}

While the concepts from above allow handling bit-flip errors (something that might have been sufficient in the classical world), the quantum world allows for a substantially larger continuum of errors. For example, recall the depolarization error introduced in Section~\ref{sec:noise_in_quantum_computing}, which sets a qubit to a completely random state---this cannot be corrected using bit-flip correction only.
Even worse, qubits also contain phase information, which has no classical counterpart at all and which must also be protected. 
Fortunately, it turns out that it is sufficient to correct only a subset of errors to correct \emph{all} possible (unitary) errors that can occur in quantum computers~\cite{Knill1996}.
In practice, it suffices to consider bit-flip and phase-flip errors only to have a universal \mbox{error-correcting} code~\cite{Roffe2019}.
In 1995 Peter Shor developed the first quantum \mbox{error-correcting} code that accomplished just that~\cite{shor1995}. Since then, multiple types of quantum \mbox{error-correcting} schemes have been developed such as CSS codes~(\cite{shor1995,Steane1996,Calderbank1996}), stabilizer codes~(\cite{Gottesman1996}), surface codes~\cite{Bravyi1998,Dennis2002}, and adaptive \mbox{error-correcting} codes optimized for specific quantum hardware~\cite{Fletcher2008,Robertson2017}.

\begin{figure}
		\centering
		\begin{tikzpicture}
		\begin{axis}[
            legend style={at={(1,1)}, anchor=north east, fill=none, draw=none},
            nodes={scale=0.8, transform shape},		  
            line width=1pt,			
            grid=both,
            width=7cm,		
			ylabel near ticks, 
			xlabel near ticks,                        
            grid style={line width=.1pt, draw=gray!30},
            major grid style={line width=.2pt,draw=gray!50},	
            scaled x ticks = false,
            label style={font=\normalsize},
			yticklabel style={
       			/pgf/number format/fixed,
        		/pgf/number format/precision=2
			},  
			ytick={0.5, 0.6,0.7,0.8,0.9,1},
			ylabel=fidelity,             
            xticklabel style={
       			/pgf/number format/fixed,
        		/pgf/number format/precision=2
			}, 
			xtick={0, 0.1,0.2,0.3,0.4,0.5},
			xlabel=error probability,
            ]
            \addplot[smooth, color=red!50!gray] table[col sep=comma, x=ErrorProb, y=NoECC] {./pgfplot/Shor3Noise.csv};
            \addlegendentry{without error correction}
            \addplot[smooth, color=blue!50!gray] table[col sep=comma, x=ErrorProb, y=Bit-flipECC] {./pgfplot/Shor3Noise.csv};
            \addlegendentry{ideal bit-flip correction}
            \addplot[smooth, color=brown!50!gray] table[col sep=comma, x=ErrorProb, y=Bit-flipECCRealistic] {./pgfplot/Shor3Noise.csv};
            \addlegendentry{realistic bit-flip correction}
		\end{axis}
	\end{tikzpicture}	
	\vspace*{-3mm}	
	\caption{State fidelity with increasing error probability}
	\label{shor3_eval}
	\vspace*{-6mm}
\end{figure}
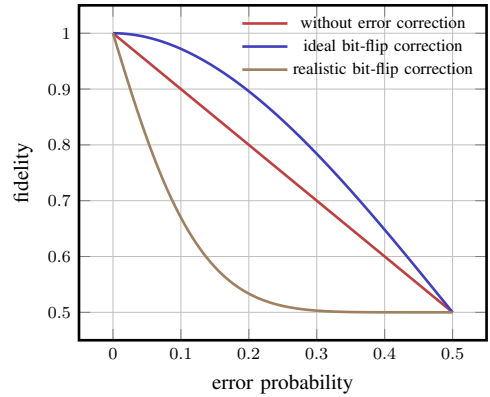

\subsection{Motivation}

\begin{figure*}
	\begin{subfigure}[b]{0.2\linewidth}
	\begin{quantikz}
	\lstick{$\ket{0}$} 		 & \gate{X} & \qw \\
						     &			&	  \\
						     &			&	  \\	
						     &			&	  \\
						     &			&	  \\					     
	\end{quantikz}
	\caption{Original circuit}
	\label{fig:simple_circuit}	
	\end{subfigure}\hfill
	\begin{subfigure}[b]{0.8\linewidth}
	\includegraphics[width=14cm]{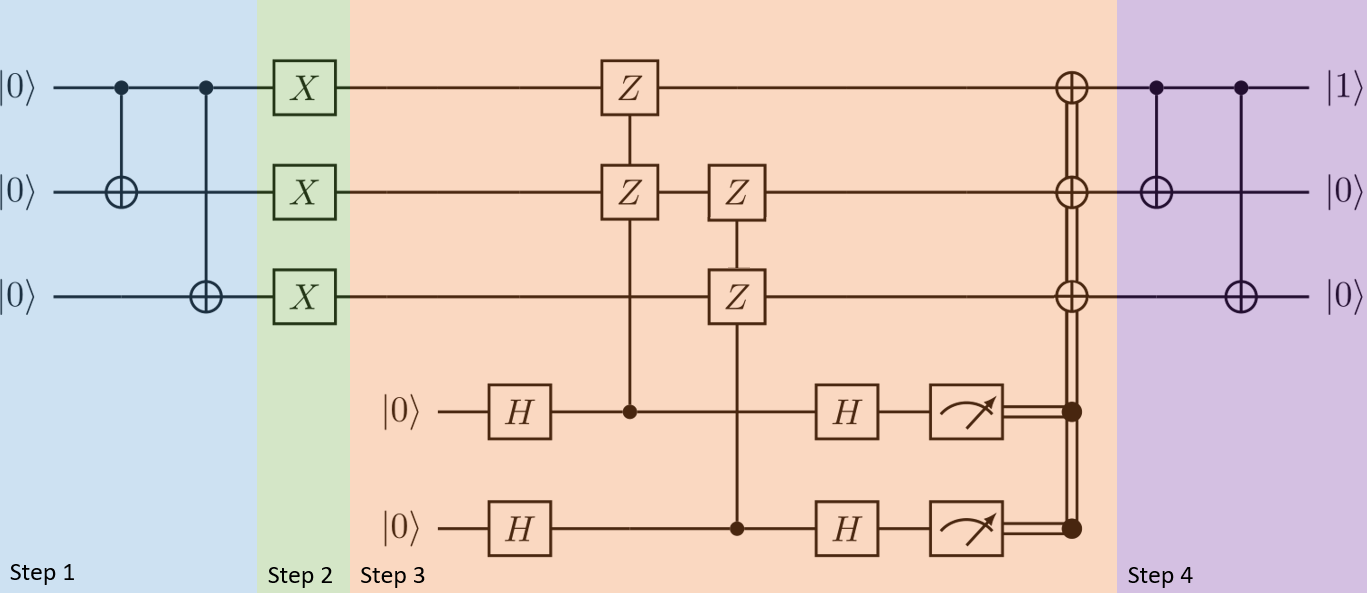}
	\caption{Circuit with bit-flip error correction}
	\label{fig:complex_circuit}	
	\end{subfigure}\hfill
	\vspace*{-3mm}
	\caption{Applying the bit-flip \mbox{error-correcting} code to a circuit.}
	\label{fig:simple_circuit_with_shor3_ecc}
	\vspace*{-3mm}
\end{figure*}

As shown above, quantum error correction addresses the problem of errors within quantum computations and there is a multitude of error correction schemes available. However, using these schemes is not as straightforward as Example~\ref{exp:shor3} suggests. 
Noisiness is not restricted to specific erroneous quantum channels but affects \emph{all} operations (including the operations necessary for error correction itself).

\begin{example}
\label{exp:shor3Analysis}
The effects of noise are different depending on the simulated hardware model. To illustrate this, in Fig.~\ref{shor3_eval}, we plotted the fidelity\footnote{The fidelity is a metric that expresses the overlap between two states~\cite{Jozsa1994}. Here we use it to express the overlap between the state with and without errors.} of state $\ket{0}$ depending on the probability of a bit-flip error in three different scenarios---without any kind of error protection (red curve), with the ideal bit-flip error protection presented in Example~\ref{exp:shor3} (blue curve), and with a more realistic bit-flip error protection (brown curve). In the latter case, we assumed that the operations necessary to realize the bit-flip error correction are also affected by noise (this is reasonable to assume, as discussed in Section~\ref{sec:noise_in_quantum_computing}).
Without any kind of error correction, the fidelity of the encoded state more or less linearly declines with increasing error probability. Using the ideal bit-flip error protection increases the fidelity, as long as the error probability stays below 50~\%. Finally, in the realistic case the fidelity severely drops compared to the case where no error correction is applied at all. 
\end{example}

As illustrated by the example, naively applying error correction schemes to quantum circuits may even lower the fidelity (as proven by the threshold theorem~\cite{knill1998,Aharonov1997,Kitaev1997}). The optimal error correction scheme depends on the properties of the simulated circuit as well as the considered quantum computer. Additionally, it is necessary to optimally tailor the error correction scheme to the specific use case. Thus
a thorough evaluation is needed.

However, while there exists a lot of research on quantum error correction (e.g., \cite{shor1995,Steane1996,Calderbank1996,Gottesman1996,Bravyi1998,Dennis2002,Fletcher2008,Robertson2017}), the focus clearly (and understandably) is on the development of corresponding concepts for \mbox{error-correcting} codes and proof-of-concept implementations on selected quantum circuits. 
As illustrated above, however, evaluating \mbox{error-correcting} codes requires a broader consideration---thus far, involving extensive research and tedious manual implementation of said schemes. To address this problem, we propose a framework that automates the process of applying error correction to circuits and also allows for noise-aware quantum circuit simulation.

\section{Proposed Framework}
\label{sec:framework}

The proposed framework is supposed to support the process of evaluating error correction schemes. To this end, it does not only allows to apply \mbox{error-correcting} codes to a given quantum circuit but also supports the entire process of circuit preparation and simulation. More precisely, the framework (1)~automates the circuit compilation flow (including the application of \mbox{error-correcting} codes), (2) provides quantum circuit simulation with noise-aware quantum circuit simulators, and (3) is flexible and easily extensible for new use cases. 

Note that, while circuit compilation and simulation are necessary parts of such a framework, there already exist available solutions (e.g.,~\cite{gidney2021stim,qiskit,grurl23,cirqFramework2022,guerreschiIntelQuantumSimulator2020}). Therefore, the proposed framework is realized in such a manner that it can be coupled with these solutions. %
In the remainder of this section, the proposed framework is presented.

\subsection{Compiler}
\label{sec:compiler}

During the compilation process, the quantum circuit is prepared for the circuit simulation or execution on a real quantum computer. 
Here, we focus on the main functionality that is provided by the framework, namely the automatic application of \mbox{error-correcting} codes to the quantum circuit. 
For this, we exploit the fact that, while there exist many different error correction schemes, all of them must in practice realize the same steps when applied to a quantum circuit. These steps are:

\subsubsection*{1. Qubit Encoding} 
The basic idea of all \mbox{error-correcting} codes is that information is encoded in such a way that it allows detection and correction of occurring errors. While encoding depends on the respective approach, all schemes have in common that information is distributed among multiple qubits. At the end of this step, all qubits within the logical circuit are encoded into logical qubits consisting of multiple physical qubits.

\subsubsection*{2. Operation Encoding} 
Having a quantum circuit with encoded qubits, the next step accordingly adjusts the operations. More precisely, the originally intended operations are mapped to logical operations, i.e., operations that are functionally equivalent but consider the applied qubit encoding.
How each operation is mapped is not always trivial~\cite{eastin2009} and depends on the used error correction scheme, which is why only specific quantum operations are supported for each \mbox{error-correcting} code.

\subsubsection*{3. Error Correction} 
The error detection and correction routines are added to the circuit. This consists of syndrome extraction %
followed by some corrective operations on the logical qubits---depending on the extracted syndrome.
The optimal frequency of executing the detection and correction routine, i.e., after how many quantum operations this routine is applied to a logical qubit, depends on several factors, e.g., the hardware specifications of the designated quantum computer or the error-proneness of the application. To accommodate this, the frequency of the detection and correction step can be freely adjusted.

\subsubsection*{4. Qubit Decoding} Finally, in the last step, the logical qubit is decoded back to a single physical qubit, so that it can be measured. For convenience, this is realized in such a way, that the ordering of the output qubits is equal to the order of the qubits in the original circuit.

The following example illustrates the process of applying an \mbox{error-correcting} code using the bit-flip code presented in Example~\ref{exp:shor3}.

\begin{example}
\label{exp:encoding}
In Fig.~\ref{fig:simple_circuit_with_shor3_ecc}, a circuit without (Fig.~\ref{fig:simple_circuit}) and with (Fig.~\ref{fig:complex_circuit}) bit-flip error correction is presented. 
The color-coding in the circuit with error correction represents during which step the respective parts are generated.
During the encoding, each qubit of the original circuit is encoded into a logical qubit by entangling it with two ancillary qubits. %
Next, each quantum operation is copied for each qubit making up the logical qubit, e.g., an $\operatorname{X}$ operation in the original circuit is mapped to an $\operatorname{X}$ operation onto each qubit making up the respective logical qubit. Afterwards, error detection and correction is added. For the error detection, a projective measurement is added, which is realized by two ancillary qubits and four controlled $\operatorname{Z}$ operations. The ancillary qubits then contain the syndrome, which indicates whether a bit-flip error has occurred. If that is the case, the flipped qubit is corrected by applying an $\operatorname{X}$ operation to it. Finally, the qubit is decoded so that it can be measured. 
\end{example}

Since these steps are necessary for all \mbox{error-correcting} codes, the framework provides templates for realizing them.
Due to this modular design, the framework can easily be extended, either by modifying already available \mbox{error-correcting} codes or by adding new ones. 
Currently the framework supports four \mbox{error-correcting} codes ranging in size to different error correction types, namely %
the Shor Code~\cite{shor1995} (as this is the first code that corrects arbitrary single-qubit errors), the Laflamme Code~\cite{Laflamme1996} (as this is the smallest possible code protecting against arbitrary single-qubit errors), the Steane code~\cite{Steane1996} (as it is a well-studied code used in other experiments such as~\cite{Robertson2017}), and a surface code~\cite{Wootton2017,Fowler2009} (due to their popularity in the current research on \mbox{error-correcting} codes). 

\subsection{Simulator}
\label{sec:simulator}

An essential part of the framework is the ability to directly simulate circuits with (and without) error correction and different hardware models. 
To this end, several simulation approaches have recently been introduced in the literature (e.g.,~\cite{gidney2021stim,qiskit,grurl23,cirqFramework2022,guerreschiIntelQuantumSimulator2020}). Some of these are available as \mbox{open-source}, and hence can be readily integrated into the proposed framework. 
By this, we gain access to multiple simulation styles, which come with their advantages/drawbacks and the most appropriate simulator can be selected for each use case.

\begin{figure}
	\centering
		\begin{tikzpicture}
		\begin{axis}[
            legend style={at={(0.00,1)}, anchor=north west, fill=none, draw=none},
            nodes={scale=0.8, transform shape},		  
			line width=1.5pt,
			width=7cm,				
			grid=both,
    		grid style={line width=.1pt, draw=gray!30},
            major grid style={line width=.2pt,draw=gray!50},
            ylabel=runtime in seconds,  
			xlabel=number of simulated qubits,   	
			]
			\addplot[smooth, blue!50!gray] table[col sep=comma, x=qubit, y=density] {./pgfplot/simulatorComparison.csv};
			\addlegendentry{density matrix simulator\hspace*{1mm}}
			\addplot[smooth, orange!50!gray] table[col sep=comma, x=qubit, y=stoch] {./pgfplot/simulatorComparison.csv};
			\addlegendentry{stochastic array-based simulator\hspace*{5mm}}
			\addplot[smooth, green!50!gray] table[col sep=comma, x=qubit, y=stab] {./pgfplot/simulatorComparison.csv};
			\addlegendentry{stabilizer-based simulator}	
		\end{axis}
	\end{tikzpicture}
	\vspace*{-3mm}
	\caption{Runtime of noise-aware quantum circuit simulation depending on the simulation style}
	\label{fig:comparingSimulationStyles}
	\vspace*{-6mm}
\end{figure}
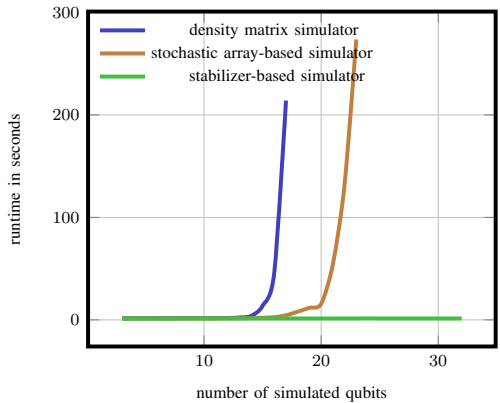

To illustrate these differences, we simulated the entanglement circuits (which construct the GHZ state over all qubits) with an increasing number of qubits. During all simulations, we applied depolarization noise with 0.001~\% probability to the qubit whenever it has been used. We ran this experiment with Qiskit~\cite{qiskit} using three different simulation styles, namely, a density matrix simulator, a stochastic array-based simulator (with 2000 shots), and a stabilizer-based simulator (with 2000 shots). In Fig.~\ref{fig:comparingSimulationStyles} the simulation times are plotted as a function of the number of simulated qubits.
The plot shows that the runtime of the density matrix simulator and the stochastic array-based simulator grows exponentially with increasing qubits, while the stabilizer-based simulator exhibits a polynomial runtime. However, while the density matrix simulator performs worst in runtime, it allows to simulate arbitrary quantum operations and gives a \emph{full} description of the final state. The stochastic array-based simulator also allows to simulate arbitrary quantum operations but only approximates the final quantum state. In contrast, the stabilizer-based simulator is restricted to a non-universal gate set of quantum operations (i.e., the Clifford gate set).  

This small experiment shows the advantage of having a variety of different quantum circuit simulators available for experiments. Currently Qiskit~\cite{qiskit} is used for quantum circuit simulation. We chose Qiskit as it offers not only different noise-aware quantum circuit simulators but also access to real quantum hardware. Naturally, the quantum circuit simulator can easily be switched to different software.

\section{Application and Demonstration}
\label{sec:evaluation}

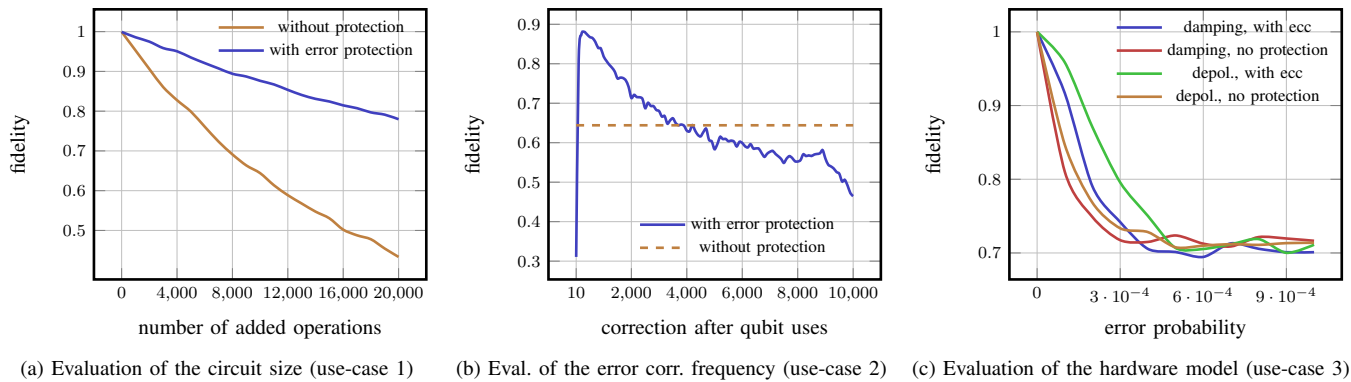
\begin{figure*}[t]
	\begin{subfigure}[b]{0.33\linewidth}
		\centering
				\begin{tikzpicture}
				\begin{axis}[
				legend style={at={(1,1)}, anchor=north east, fill=none, draw=none},
				nodes={scale=0.8, transform shape},	
				line width=1pt,			
				grid=both,
				width=6cm,
    			grid style={line width=.1pt, draw=gray!30},
    			major grid style={line width=.2pt,draw=gray!50},
    			x tick label style={/pgf/number format/fixed},    				
            	scaled x ticks = false,
            	label style={font=\normalsize},
				yticklabel style={
       				/pgf/number format/fixed,
        			/pgf/number format/precision=2
				},  
				ytick={0.5, 0.6,0.7,0.8,0.9,1},
				ylabel=fidelity,            
            	xticklabel style={
       				/pgf/number format/fixed,
        			/pgf/number format/precision=2
				}, 
				xtick={0, 4000, 8000, 12000, 16000, 20000},
				xlabel=number of added operations,  			
				]
				\addplot[smooth, color=orange!50!gray] table[col sep=comma, x=depth, y=noProtection] {./pgfplot/gateDepthBigEntFQ500_0.02.csv};
				\addlegendentry{without protection}
				\addplot[smooth, color=blue!50!gray] table[col sep=comma, x=depth, y=withProtection] {./pgfplot/gateDepthBigEntFQ500_0.02.csv};
				\addlegendentry{with error protection}
				\end{axis}
				\end{tikzpicture}	
		\caption{Evaluation of the circuit size (use-case 1)}
		\label{fig:gate_depth}
	\end{subfigure}\hfill
	\begin{subfigure}[b]{0.33\linewidth}
		\centering
				\begin{tikzpicture}
				\begin{axis}[
				legend style={at={(0.90,0.05)},anchor=south east, fill=none, draw=none},
				nodes={scale=0.8, transform shape},	
				line width=1pt,			
				grid=both,
				width=6cm,
    			grid style={line width=.1pt, draw=gray!30},
    			major grid style={line width=.2pt,draw=gray!50},
    			x tick label style={/pgf/number format/fixed},    				
            	scaled x ticks = false,
            	label style={font=\normalsize},
				yticklabel style={
       				/pgf/number format/fixed,
        			/pgf/number format/precision=2
				},  
				ytick={0.3, 0.4, 0.5, 0.6,0.7,0.8,0.9,1},
				ylabel=fidelity,            
            	xticklabel style={
       				/pgf/number format/fixed,
        			/pgf/number format/precision=3
				}, 
				xtick={10, 2000, 4000, 6000, 8000, 10000},
				xlabel=correction after qubit uses,  			
				]
				\addplot[smooth, color=blue!50!gray] table[col sep=comma, x=fq, y=5] {./pgfplot/gateDepthBigEntFQ.csv};
				\addlegendentry{with error protection}
				\addplot[dashed, orange!50!gray] table[col sep=comma, x=fq, y=noecc] {./pgfplot/gateDepthBigEntFQ.csv};
				\addlegendentry{without protection}
				\end{axis}
				\end{tikzpicture}	
		\caption{Eval. of the error corr. frequency (use-case 2)}
		\label{fig:ecc_fq}		
	\end{subfigure}	
		\begin{subfigure}[b]{0.33\linewidth}
		\centering
				\begin{tikzpicture}
				\begin{axis}[
				legend style={at={(1,1)}, anchor=north east, fill=none, draw=none},
				nodes={scale=0.8, transform shape},	
				line width=1pt,			
				grid=both,
				width=6cm,
    			grid style={line width=.1pt, draw=gray!30},
    			major grid style={line width=.2pt,draw=gray!50},
            	scaled x ticks = false,
            	label style={font=\normalsize},
				yticklabel style={
       				/pgf/number format/fixed,
        			/pgf/number format/precision=2
				},  
				ytick={0.5, 0.6,0.7,0.8,0.9,1},
				ylabel=fidelity,            
				xtick={0, 0.0003, 0.0006, 0.0009},
				xlabel=error probability,  			
				]
				\addplot[smooth, color=blue!50!gray] table[col sep=comma, x=error, y=SteaneQ] {./pgfplot/hwModel10000EntFQ500.csv};
				\addlegendentry{damping, with ecc}
				\addplot[smooth, color=red!50!gray] table[col sep=comma, x=error, y=Q] {./pgfplot/hwModel10000EntFQ500.csv};
				\addlegendentry{damping, no protection}
				\addplot[smooth, green!50!gray] table[col sep=comma, x=error, y=SteaneD] {./pgfplot/hwModel10000EntFQ500.csv};
				\addlegendentry{depol., with ecc}
				\addplot[smooth, orange!50!gray] table[col sep=comma, x=error, y=D] {./pgfplot/hwModel10000EntFQ500.csv};
				\addlegendentry{depol., no protection}
				\end{axis}
				\end{tikzpicture}		
		\caption{Evaluation of the hardware model (use-case 3)}
		\label{fig:hardware_modell}
	\end{subfigure}	
	\vspace*{-4mm}
	\caption{Experiments}
	\label{fig:experiments}
	\vspace*{-6mm}
\end{figure*}

We demonstrate the usefulness of the proposed framework by considering the Steane code~\cite{Steane1996} as a well-known representative of an \mbox{error-correcting} code. For this code, we evaluated in which scenarios the error correction actually improves the reliability and in which it does not\footnote{Using the proposed framework, similar considerations can be done with other codes as well.}.
To this end, we evaluated the effects of three different properties:
\begin{enumerate}
\item Size of the considered circuit: error correction is a costly procedure (see Fig.~\ref{fig:simple_circuit_with_shor3_ecc}). Whether error correction improves the reliability of a circuit substantially depends on its size.
	\item Frequency of error correction: Detecting and correcting errors is a costly operation. The proper frequency with which this routine is performed is critical when applying an \mbox{error-correcting} code. 
	\item Assumed hardware model: The considered hardware model (i.e., error types and error probability) also effects the usefulness of the quantum circuit. For example, if the error probability is too high, the additional operations introduced by error correction may actually degrade the quality of the results.
\end{enumerate}

For the evaluation, we considered a benchmark that creates the GHZ state between all qubits (taken from~\cite{quetschlichMQTBenchBenchmarking2022a}). We chose this benchmark because its simplicity makes it the ideal candidate to test the Steane code in different scenarios, without having to account for fragments within the data. Besides that, the benchmark can be easily varied in size, and the created states are strongly correlated thereby making (uncorrected) errors easily noticeable.
If not stated otherwise, the following default parameters have been used in the evaluation:
\begin{itemize}
\item Depolarization noise (applied whenever a qubit is used) has been considered with an error probability of 0.001~\%.
\item Error correction and detection has been applied to a logical qubit prior to the measurement and whenever it has been used 500 times.
\item The entanglement benchmark is used with five qubits and 10.000 dummy operations added to the end of the circuit (the dummy operations do not change the result but are affected by noise).
\end{itemize}

For each property discussed above, this default use-case is adapted to evaluate the respective property.
That is, for (1) we simulated entanglement circuits with an increasing number of dummy operations, for (2) we simulated entanglement circuits with a decreasing rate of error correction steps, and, finally, for (3) we simulated the benchmark with increasing error probability not only with depolarization noise (mimicking gate errors) but also with amplitude damping errors (mimicking coherence errors). In the latter case we used the entanglement benchmark with two qubits (instead of five), as we had to use a stochastic array-based simulator for simulating the amplitude damping noise.

We used the proposed framework to implement the Steane code to the circuits for all runs and simulated the circuits using Qiskit~\cite{qiskit}---which, as described in Section~\ref{sec:simulator}, has been integrated into the framework. More precisely, we used the stabilizer-based simulator when considering depolarization errors and the state vector simulator when considering amplitude damping errors (as these cannot be simulated with the stabilizer simulator). We simulated all experiments stochastically with 2000 shots. The simulations yield a probability distribution for measuring specific basis states, which we use to calculate the ``classical'' fidelity (also known as the Hellinger coefficient~\cite{Bhattacharyya1946}) between the expected probability distribution (i.e, the received probability distribution when no errors occur during the computation) and the actual distribution received by the noise-aware quantum circuit simulation. 
We chose this distance measure over others, such as the fidelity between the quantum states~\cite{Jozsa1994} or the average fidelity (as used for example in~\cite{gilchrist2005}), as we want to evaluate how a noisy quantum computer %
affects the ``correctness'' of the measured final states for \emph{specific} quantum algorithms.
For these evaluations, we conducted 285 simulations and applied error correction to the quantum circuits 137 times.

Fig.~\ref{fig:experiments} summarizes the results of the experiments.
Fig~\ref{fig:gate_depth} depicts the fidelity with increasing depth of the entanglement circuits.  It shows that the usefulness of error correction increases with the gate depth. Conducting error correction only pays off when the circuit is of sufficient depth. 
Next, in Fig.~\ref{fig:ecc_fq} the fidelity is plotted with decreasing error correction frequency. Contrary to intuition, it is not advantageous to correct errors as often as possible during the computation. This can be explained by the fact that, as discussed above, error correction is expensive and adds a substantial amount of extra operations. Hence, it has a positive effect if error correction application and resulting overhead is traded-off (a task for which the proposed framework offers a very helpful tool).
Finally, Fig.~\ref{fig:hardware_modell} depicts the fidelity with increasing (physical) error probability and simulated error type. The plot shows that error correction is only useful, as long as the error probability stays low. Additionally, the simulated noise type noticeably affects the effectiveness of the error correction scheme. Thus, depolarization errors could be handled more effectively compared to amplitude damping errors.

Obviously, these demonstrations only provide a snippet of what kind of evaluations can be conducted with the proposed framework. But it clearly demonstrates the usefulness of the approach. While, until now, it required a lot of tedious manual work to apply error correction to circuits and evaluate the corresponding effects, now the proposed framework automates the task---making quantum error correction more broadly applicable and easier to analyze.

\section{Conclusions}
\label{sec:conclusions}

Quantum \mbox{error-correcting} codes are an essential part of building scalable and resilient quantum hardware. However, currently most of the corresponding work in this domain heavily relies on manual labor and/or is based on theoretical results only. We address this problem by proposing an \mbox{open-source} framework that automates the process of applying error correction and also allows noise-aware simulation of the protected circuits. We demonstrate the advantages and usefulness of the proposed framework by evaluating the reliability of the Steane \mbox{error-correcting} code with respect to different parameters. The proposed framework is published as \mbox{open-source} (available at~\href{www.github.com/cda-tum/qecc}{www.github.com/cda-tum/qecc}) and implemented in a modular fashion, so that it can be easily configured and extended for new \mbox{error-correcting} codes and to make quantum error correction broadly applicable.

\section*{Acknowledgments}
This work received funding from the University of Applied Sciences PhD program of the State of Upper Austria (managed by the FFG), from the European Research Council (ERC) under the European Union’s Horizon 2020 research and innovation program (grant agreement No. 101001318), was part of the Munich Quantum Valley, which is supported by the Bavarian state government with funds from the Hightech Agenda Bayern Plus, and has been supported by the BMWK on the basis of a decision by the German Bundestag through project QuaST.

\printbibliography
\end{document}